# Development of novel ionization chambers for reference dosimetry in electron FLASH radiotherapy

Kevin Liu[a,b] | Shannon Holmes[c] | Ahtesham Ullah Khan[d,e]| Brian Hooten[c] | Larry DeWerd[c] | Emil Schüler[a,b†*] | Sam Beddar[a,b†*]

[a] Department of Radiation Physics, Division of Radiation Oncology, The University of Texas MD Anderson Cancer Center, Houston, Texas, USA

[b] The University of Texas MD Anderson Cancer Center UTHealth Houston Graduate School of Biomedical Sciences, Houston, Texas, USA

[c] Standard Imaging Inc., Middleton, WI, USA

[d] Department of Medical Physics, School of Medicine and Public Health, University of Wisconsin-Madison

[e] Department of Radiation Oncology, Northwestern Memorial Hospital, Northwestern University Feinberg School of Medicine, Chicago, IL, USA

*Co-senior/co-corresponding authors:

Emil Schüler, PhD, and Sam Beddar, PhD

Department of Radiation Physics, Division of Radiation Oncology

The University of Texas MD Anderson Cancer Center

1515 Holcombe Blvd, Houston, TX 77030, USA.

Emails: eschueler@mdanderson.org, abeddar@mdanderson.org

†Both senior authors contributed equally

ABSTRACT

**Background:** Reference dosimetry in ultra-high dose rate (UHDR) beamlines is significantly hindered by limitations in conventional ionization chamber design. In particular, conventional chambers suffer from severe charge collection efficiency (CCE) degradation in high dose per pulse (DPP) beams.

**Purpose:** The aim of this study was to optimize the design and performance of parallel plate ion chambers for use in UHDR dosimetry applications, and evaluate their potential as reference class chambers for calibration purposes. Three chamber designs were produced to determine the influence of the ion chamber response on electrode separation, field strength, and collection volume on the ion chamber response under UHDR and ultra-high dose per pulse (UHDPP) conditions.

**Methods:** Three chambers were designed and produced: the A11-VAR (0.2-1.0 mm electrode gap, 20 mm diameter collector), the A11-TPP (0.3 mm electrode gap, 20 mm diameter collector), and the A30 (0.3 mm electrode gap, 5.4 mm diameter collector).The chambers underwent full characterization using an UHDR 9 MeV electron beam with individually varied beam parameters of pulse repetition frequency (PRF, 10-120Hz), pulse width (PW, 0.5-4us), and pulse amplitude (0.01-9 Gy/pulse). The response of the ion chambers was evaluated as a function of the DPP, PRF, PW, dose rate, electric field strength, and electrode gap.

**Results:** The chamber response was found to be dependent on DPP and PW, whose dependencies were mitigated with larger electric field strengths and smaller electrode spacing. At a constant electric field strength, we measured a larger CCE as a function of DPP for ion chambers with a smaller electrode gap in the A11-VAR. For ion chambers with identical electrode gap (A11-TPP and A30), higher electric field strengths were found to yield better CCE at higher DPP. A PW dependence was observed at low electric field strengths (500 V/mm) for DPP values ranging from 1-5 Gy at PWs ranging from 0.5-4 μs, but at electric field strengths of 1000 V/mm and higher, these effects become negligible.

**Conclusions:**

This study confirmed that the charge collection efficiency of ion chambers depends strongly on the electrode spacing and the electric field strength, and also on the DPP and the PW of the

UHDR beam. The new finding of this study is that the pulse width (PW) dependence becomes negligible with reduced electrode spacing and increased electric field. A CCE of ≥ 95% was achieved for DPPs of up to 5 Gy with no observable dependence on PW using the A30 chamber, while still achieving an acceptable performance in conventional dose rate beams, opening up the possibility for this type of chamber to be used as a reference class chamber for calibration purposes of electron FLASH beamlines.

## 1 | INTRODUCTION

Radiation therapy (RT) has long been a cornerstone in the treatment and management of various cancers, with crucial roles in tumor control and palliation. Technological advancements in radiation delivery systems over time have led to significant improvements in treatment precision and reduced collateral damage to healthy tissues[1]. FLASH RT could be the next phase of such advancements, in which radiation is delivered at ultra-high dose rates (UHDR) and ultra-high dose per pulse (UHDPP) conditions[2-5]. This emerging modality allows a typical radiation treatment to be completed in milliseconds instead of minutes, providing significant potential for improving treatment outcomes through improved precision by negating the effects of intra-fraction tumor motion during beam delivery. Preclinical and clinical studies have shown promising results in many different organ systems, and in multiple animal species, related to sparing normal tissue with maintained tumor control.[2,6-12] However, variations in the magnitude of the FLASH effect have been reported, with some studies reporting no sparing effects under UHDR conditions relative to conventional (CONV) dose rate irradiations[13-15]. A likely factor contributing to variations in the FLASH effect between different studies arises from inconsistencies in the physical beam parameters used in UHDR beamlines arising from limitations and uncertainties related to obtaining accurate dosimetry under UHDR conditions[3].

Currently, dosimeters heavily relied upon for measurements in UHDR beamlines are alanine, thermoluminescent dosimeters (TLDs), optically stimulated luminescent dosimeters (OSLDs), and radiochromic film, owing to their dose-rate independence[16-21]. Thus institutions that use these dosimeters have adopted protocols for *in vivo* studies that involve multiple dose rate-independent dosimeters with different physical mechanisms of signal generation, and thereby different dependencies on radiation, for concurrent and redundant dosimetry in UHDR beamlines.[22-24] Drawbacks to using these dosimeters are their lack of temporal resolution (passive detectors), delayed read-out, and large uncertainties compared with conventional ion chambers used in routine clinical practice. These drawbacks have prompted some to instead rely on alternative detectors as diodes, diamonds, Faraday cups, plastic scintillators, and beam current transformers (BCTs), which have the potential to give time-resolved dosimetric readouts but are still limited by radiation damage and the need for consistent re-calibration to the aforementioned dose rate-independent dosimeters to monitor beam output in real time, with corresponding propagation of the uncertainty inherent in these passive dosimeters.[25-31]

Although FLASH RT offers immense potential, its UHDR and UHDPP conditions pose unique challenges to conventional dosimeters capable of real-time measurement, most importantly ion chambers,[28] which are recognized by national and international protocols for reference dosimetry in RT.[32,33] The ultra-rapid delivery of radiation has been reported to lead to saturation effects, caused by high charge recombination losses (in existing ion chambers) and electronic saturation (in existing electrometer systems) that cannot be accounted for by using established correction factors.[34-36] Consequently, such commercially available ion chambers are unable to accurately measure the actual dose delivered in UHDR and UHDPP beamlines, hindering their use and the clinical translation of FLASH RT.[27] Translating FLASH RT to clinical practice requires bringing dosimetric reporting up to the standards used for conventional dose-rate deliveries[33] in having the capability to calibrate FLASH-capable units using established protocols.[28]

To address the limitations of ion chambers, researchers have investigated the response of various commercially available ion chambers in UHDR and UHDPP beamlines and have published novel designs that are constructed or simulated for ion chambers.[37-41] DPP was found recently to be the main beam parameter affecting chamber response, and other beam parameters such as pulse repetition frequency (PRF) and pulse width (PW), which influence mean and instantaneous dose rate, respectively, have lesser effects on the chambers investigated and the beam parameters tested.[37,40,42] Smaller electrode spacing and higher electric field strengths (corresponding to higher bias voltages) in the ion chamber design were found to be promising strategies to mitigate the effects of ion recombination by allowing faster charge collection.[38-40,42] However, such tasks are not trivial, as reproducibly manufacturing ion chambers with sub-mm electrode spacings and using them with high electric field gradients (> 300 V/mm) are not common practice in RT dosimetry. This study aimed to evaluate the performance and viability of next-generation air-vented parallel-plate ion chambers with ultra-thin (sub-mm) electrode spacing under UHDR beamline conditions, with the goal of demonstrating their suitability and use for reference dosimetry applications in UHDR conditions.

## 2 | METHODS

### 2.1.1 | Chamber design

The prototype ion chambers evaluated in this study were based on 1) the Exradin A11 Parallel Plate Ion Chamber (Standard Imaging) modified to allow for a variable electrode spacing ranging from 0.2 mm to 1.0 mm (the Exradin A11-VAR); 2) the Exradin A11 Parallel Plate Ion Chamber (Standard Imaging) modified to have a fixed electrode spacing of 0.3 mm (the Exradin A11-TPP); and 3) the Exradin A10 Parallel Plate Ion Chamber (Standard Imaging, Middleton, WI, USA) modified to have a electrode spacing of 0.3 mm (the Exradin A30). The A11-VAR included a base ion chamber that had different entry windows that could be manually changed to implement the desired gap thickness provided by the manufacturer (0.2-1.0 mm). The goal of this study was to evaluate ion chambers designed with sub-mm electrode spacing but with subtle differences in their design features to compare their performance in UHDR electron beamlines at different electric field strengths. The relevant specifications for each ion chamber used are shown in Figure 1 and listed in Table 1.

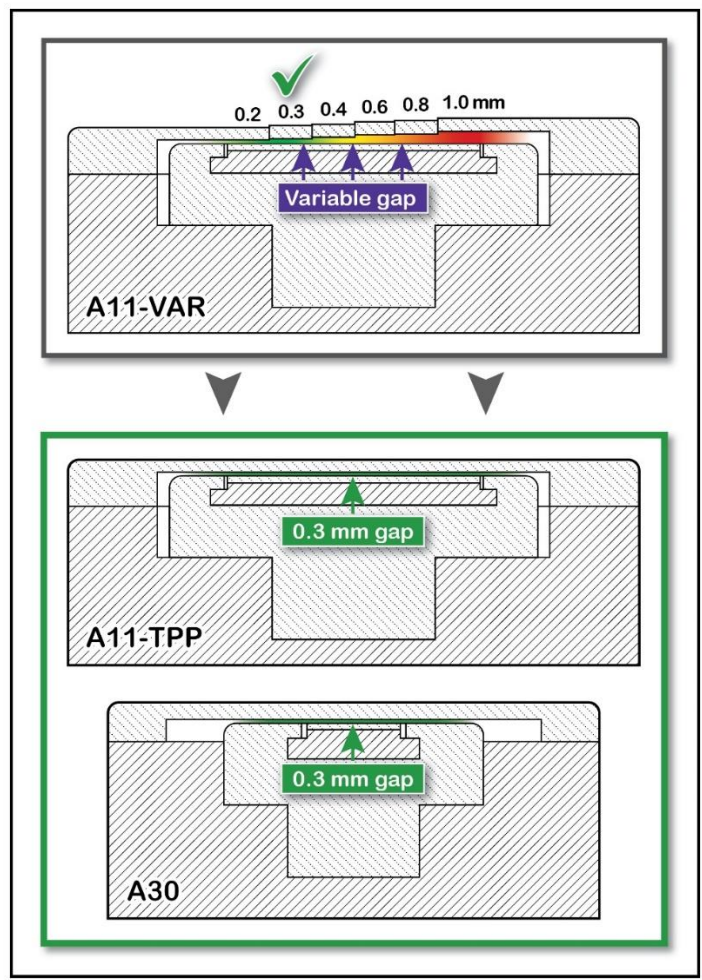

**Figure 1.** Schematics of the A11-VAR showing different gap thicknesses, the A11-TPP, and A30 ion chambers.

TABLE 1  Parallel plate ion chamber specifications

|  | Parallel Plate Ion Chambers Tested | | |
| --- | --- | --- | --- |
|  | A11-VAR | A11-TPP | A30 |
| **Chamber type** | Parallel plate | Parallel plate | Parallel plate |
| **Volume, cm³** | 0.062–0.31 | 0.093 | 0.0075 |
| **Collector gap, mm** | 0.2–1.0 | 0.3 | 0.3 |
| **Collector diameter, mm** | 20 | 20 | 5.4 |
| **Guard ring width, mm** | 4.4 | 4.4 | 4.3 |

### 2.1.2 | Irradiation conditions

Before each irradiation session, the temperature and pressure were measured, recorded, and applied in the correction factors to the measured signal.[33] Each ion chamber investigated was connected to a SuperMAX electrometer (Standard Imaging), had a positive bias voltage applied to collect negative charge, had a 2-cm water-equivalent build-up cap, and had 10-cm solid water backscatter in both CONV and UHDR beamlines. The ion chambers were evaluated on the IntraOp Mobetron[43] (IntraOp, Sunnyvale, CA, USA) operating in both the FLASH and CONV modes. The Mobetron can deliver UHDR electron beams with variable parameters such as beam energy, PW, number of pulses, PRF, and the source-to-surface distance (SSD).[43] Likewise, the Mobetron can deliver a CONV dose rate 9-MeV electron beam by modifying the number of monitor units (MU) at a constant PW setting of 1.2 µs and PRF of 30 Hz. BCTs (Bergoz Instrumentation, Saint-Genis-Pouilly, France) integrated into the head of the Mobetron were used to monitor and record the beam output and parameters of individual pulses (such as PRF, PW, and pulse amplitude) delivered in UHDR conditions in real time.[25,44,45] The internal ion chamber is situated between the upper and lower BCTs and is used to monitor the number of MU delivered at CONV dose rates. For dosimetric reporting in the UHDR setting, the signal measured by the BCTs was calibrated against the dose measured with energy- and dose rate-independent radiochromic film (Gafchromic EBT3, Ashland Inc., Covington, KY, USA) at each reference setup used for the ion chambers, as reported previously.[24,25,46-48] The BCTs represent the working standard against which the ion chambers were evaluated under UHDR conditions because of their capability for evaluating UHDR beam parameters (output, PW, PRF) on a pulse-by-pulse basis. Except when explicitly noted, the setting used for UHDR on the Mobetron was a 9-MeV beam with single pulse deliveries at the measurement setup, directly below the exit window of the head of the Mobetron, to achieve the highest DPP for the PWs available.

To ascertain the chamber response at clinical dose rates and DPP values (i.e., 0.5 Gy/s and 0.015 Gy per pulse), the 9-MeV CONV dose rate beam on the IntraOp Mobetron was used. A total of 300 MU was delivered per measurement, and measurements were obtained in triplicate to calculate $P_{ion}$ and $P_{pol}$ [33] at bias voltages corresponding to the electric field strength that were evaluated for each ion chamber, which ranged from 500 V/mm to 1200 V/mm. For the 500 V/mm electric field strength, which had a corresponding bias voltage of <200 V, the $P_{ion}$ was assumed to be 1 owing to limitations in the SuperMAX electrometer in achieving bias voltages below the absolute value of 100 V. The ion chamber response under UHDR conditions was evaluated in terms of the charge collection efficiency (CCE) and $P_{pol}$ as defined in TG-51.[33] The CCE is defined as the percentage of the temperature- and pressure-corrected signal ($M_{UHDR}$) per unit dose ($D_{UHDR}$), measured under UHDR conditions by using Gafchromic film and the BCTs, to be delivered to the ion chamber in an 9-MeV UHDR electron beamline at a depth of 2 cm divided by the fully corrected signal ($M_{CONV}$) per unit dose ($D_{CONV}$), also measured with Gafchromic film, delivered to the ion chamber in a 9-MeV CONV beamline. Both irradiations were done at the exit window of the Mobetron with the ion chamber and buildup cap placed directly underneath. The CCE was calculated as shown in Equation 1.[37] The fully corrected signal includes correction factors involving $P_{ion}$, $P_{pol}$, and $P_{TP}$.[33] The $P_{TP}$ do not cancel out in Equation 1 because measurements for the ion chambers may have been obtained on different days with different temperature and pressure settings. The average ion collection times for the ion chambers were calculated from the equation: $\tau = \frac{d^2}{V \times k_2}$, where τ is the average ion collection time, d is the electrode spacing, V is the positive bias voltage setting used to collect electrons, and $k_2$ is the negative ion mobility constant, where $k_2$ = 1.89 cm$^2$/Vs[49].

$$CCE = \frac{M_{UHDR_{raw}} \times P_{TP,UHDR}/D_{UHDR}}{M_{CONV_{raw}} \times P_{TP,CONV} \times P_{ion} \times P_{pol}/D_{CONV}} \times 100\% \qquad (1)$$

When the ion chamber response is compared under different mean and instantaneous dose rate conditions, a common metric used to evaluate their response was the relative standard deviation, which is the percent ratio of the standard deviation to the mean value of the dataset evaluated. In plots of the CCE and $P_{pol}$ as a function of various beam parameters, the error bars represent two standard deviations taken from three consecutive readings and may be hidden by the measurement points because of their relatively small values.

In evaluating the ion chamber response under UHDR conditions, the A11-VAR was first characterized as a function of electric field strength under CONV conditions and DPP at 1000 V/mm for the different electrode spacings. Using the information from the A11-VAR, the A11-TPP and A30 ion chambers were later developed and subsequently characterized more comprehensively in evaluating electric field strength under CONV conditions, DPP, mean dose rate, and instantaneous dose rate at electric field strengths ranging from 500-1200 V/mm.

## 2.2 | A11-VAR

### 2.2.1 | Electric field strength

To evaluate the response of the A11-VAR for different electric field strengths at conventional dose rates, triplicate measurements were obtained at different bias voltages with 300 MU delivered. The measured signal in the A11-VAR for different electrode spacings at each electric field strength was normalized to the signal measured at an electric field strength of 1000 V/mm to compare the relative response in ion chamber signal as a function of electric field strength.

### 2.2.2 | Dose per pulse response

The dependence of the A11-VAR ion chamber response on DPP was evaluated by modifying the PW between 0.5 µs and 4 µs at a constant SSD for electrode spacings of 0.2–1.0 mm. Measurements were obtained at the highest DPP setting possible for each PW by placing the linac head directly at the surface of the buildup cap, resulting in DPP values of 1–9 Gy, corresponding to PWs of 0.5–4 µs.

## 2.3 | A11-TPP and A30

### 2.3.1 | Electric field strength

To evaluate the response of the A11-TPP and A30 ion chambers for different electric field strengths at conventional dose rates, triplicate measurements were obtained at different bias voltages with 300 MU delivered. The measured signal in the A11-TPP and A30 at each electric field strength was normalized to the signal measured at an electric field strength of 1000 V/mm to compare the relative response in the ion chamber signal as a function of electric field strength.

### 2.2.1 | Dose per pulse response

The dependence of the A11-TPP and A30 ion chamber response on DPP was evaluated by setting the PW at a constant 4 µs and changing the SSD while performing single pulse deliveries at each SSD location. The range of DPP values investigated for each chamber was 0.3–9 Gy.

## 2.4 | Mean dose rate response

The ion chamber response for the A11-TPP and A30, in terms of CCE and $P_{pol}$ as a function of mean dose rate, was evaluated by using three pulse deliveries at a DPP setting of 9 Gy and a PW setting of 4 μs. The PRF settings examined ranged from 10 Hz to 120 Hz, corresponding to mean dose rates of 135–1620 Gy/s.

## 2.5 | Instantaneous dose rate (pulse width) response at a constant dose per pulse

The instantaneous dose rate in this paper was defined as the DPP divided by the PW, which is defined at full width at half maximum (FWHM) of the pulse. The ion chamber response as a function of instantaneous dose rate was evaluated by modifying the SSD to match the output for each DPP delivered to the chambers at different PW settings, as was done previously.[50] The DPP values selected allow at least three PW settings to be evaluated with matched DPP values (DPP values evaluated were 1–5 Gy), as listed in Table 2 with their corresponding instantaneous dose rates (0.25–2.5 MGy/s).

TABLE 2. Instantaneous dose-rate values (in kGy/s) for each pulse width investigated and their corresponding doses per pulse (in Gy) ranging from 250 kGy/s to 2500 kGy/s

| Pulse Width, μs | Dose Per Pulse, Gy | | | | |
|---|---|---|---|---|---|
| | 1 | 2 | 3 | 4 | 5 |
| 0.5 | 2000 | NA | NA | NA | NA |
| 1 | 1000 | 2000 | NA | NA | NA |
| 2 | 500 | 1000 | 1500 | 2000 | 2500 |
| 3 | 333 | 667 | 1000 | 1333 | 1667 |
| 4 | 250 | 500 | 750 | 1000 | 1250 |

## 2.6 | Simulation of dose per pulse and pulse width dependence

Gotz et al. developed a theoretical model consisting of 1D time-dependent coupled partial differential equations (PDEs) to describe the recombination effects inside a plane-parallel chamber (PPC).[51] This model was used by Gomez et al. to calculate the collection efficiency of PPCs in FLASH beams as a function of electrode separation.[38] The proposed model makes the following assumptions:

1) The electric field lines are uniform in the lateral direction.
2) Charged particle drift is considered only along the direction of the electric field, and diffusion along the lateral direction is negligible. This simplifies the problem to 1D.

3) The dose distribution inside the air cavity is homogeneous.
4) The pulse separation is much greater than the pulse duration. In other words, the charged carriers created inside the air cavity are collected before the arrival of the next pulse.
5) Charge multiplication effects are ignored.

By using this model, the number density of the positive ions, $n_+$, negative ions, $n_-$, and free electrons, $n_e$, at a given time can be solved by using the following PDEs:

$$\frac{\partial n_+(x,t)}{\partial t} + \alpha n_+(x,t) n_-(x,t) + \mu_+ \frac{\partial}{\partial x}[E(x,t) n_+(x,t)] = I(x,t) \qquad (2)$$

$$\frac{\partial n_-(x,t)}{\partial t} + \alpha n_+(x,t) n_-(x,t) - \mu_- \frac{\partial}{\partial x}[E(x,t) n_-(x,t)] = \gamma_e(E) n_e(x,t) \qquad (3)$$

$$\frac{\partial n_e(x,t)}{\partial t} + \gamma_e(E) n_e(x,t) - \frac{\partial}{\partial x}[v_e(E) n_e(x,t)] = I(x,t) \qquad (4)$$

where $I(x,t)$ is the current that is produced by the radiation beam, $\alpha$ quantifies the volume recombination between the positive and negative ions, $E(x,t)$ is the electric field strength, $\mu_+/\mu_-$ describes the mobility of the ions, $\gamma_e$ denotes the attachment constant that quantifies the attachment of electrons with atoms in the air cavity to produce negative ions, and $v_e$ is the electron velocity. The presence of the heterogeneous charge carrier distribution inside the cavity and movement of ions allows the local electric field at a given location and time to be described by:

$$\frac{\partial E}{\partial x} = \frac{e}{\epsilon}[n_+(x,t) - n_-(x,t) - n_e(x,t)] \qquad (5)$$

However, because the applied potential across the electrodes is constant, the following constraint must be always met:

$$\int E(x,t) dx = V \qquad (6)$$

where $V$ is the applied bias. The number of electrons and positive ions generated by a radiation pulse can be calculated by:

$$I(x,t) = \frac{DPP \rho_{air}}{t_{pulse} e \left(\frac{W}{e}\right)_{air}} \tag{7}$$

where $DPP$ is the dose per pulse, $\rho_{air}$ is the air density, $t_{pulse}$ is the pulse duration, $e$ is the elementary charge, and $\left(\frac{W}{e}\right)_{air}$ is the mean energy required to liberate an ion pair in air. The electrons collected by the collecting electrode can be given by:

$$n_e^{collected} = \frac{\partial}{\partial x}[v_e(E) n_e(x = 0, t)] \tag{8}$$

The collection efficiency can be calculated by $(n_e^{collected} + n_-^{collected})/n_e^{generated}$. The $\alpha$, $\mu_+$, and $\mu_-$ parameters were extracted from the work of Gotz et al[51]. The $\gamma_e$ and $v_e$ parameters were extracted from the thesis work of Boissonnat as a function of the electric field strength.[52] The PDEs described in this work can be converted to ordinary differential equations (ODE) via discretization of space. The distance between the electrodes was discretized into $N$ even bins, each with size $h$. Therefore, the advection term can be re-written as:

$$\frac{\partial}{\partial x}[E(x,t) n(x,t)] \rightarrow \frac{1}{h}[E_j(x_{i+1}) n_j(x_{i+1}) - E_j(x_i) n_j(x_i)] \tag{9}$$

where $i$ is the bin index ranging from $i = 0$ to $i = N$ and $j$ is the time index ranging from $j = 0$ to $j = t$. The resulting ODE can be solved by using numerical methods. In this work, the 4$^{th}$ order Runge-Kutta method was used with adaptive step size.[53,54] The initial time step size, $t_{step}$, was chosen to be $1 \times 10^{-13}$ s and the step size was modified after each step to improve the computational efficiency of the ODE solver. This method uses the slope of the ODE to assess the magnitude of the step size. Steeper regions of the curve require small step sizes and vice versa. Implementation of the adaptive step-size method allowed computational time of <1 hour using MATLAB v2020b software (MathWorks, Natick, MA, USA). The free parameters for this model were $DPP$, $t_{pulse}$, $V$, and electrode separation. In this work, the bin size, $h$, was set to 200 nm. The simulations that were tested were intended to match and build upon the aforementioned experiments, conducted such as CCE dependence for DPP (1–10 Gy) at a constant PW of 4 μs for electric field strengths of 500-1200 V/mm. The PW dependence of ion chambers was tested for parameters of PWs ranging from 0.5-4 μs, different parameters of electric field strengths ranging from 500-1000 V/mm, and different parameters of electrode spacings ranging from 0.1-1.0 mm for DPPs ranging from 1-10 Gy.

## 3 | RESULTS

### 3.1 | A11-VAR

#### 3.1.1 Electric field strength

The mean signal (nC) measured for all the electrode spacings evaluated in the A11-VAR ion chamber (i.e., 0.2–1.0 mm) was shown to increase with increasing electric field strengths despite having approximately the same dose delivered. Relative to the signal measured at 1000 V/mm, the mean signal was found to be within ±2% at electric field strengths ranging from 100 to 1500 V/mm for the different electrode spacings tested. At electric field strengths greater than 1500 V/mm, the signal measured in the ion chambers at different electrode spacings demonstrated exponential growth in the amount of signal measured per dose delivered at CONV dose rates, indicating charge multiplication in the tested ion chamber configurations (Figure 2).

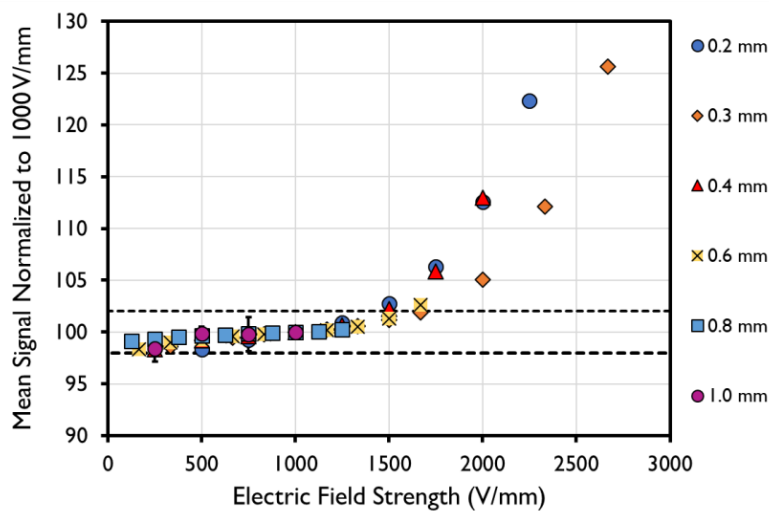

**Figure 2.** Mean signal plotted as a function of electric field strength (V/mm) normalized to the mean signal measured at 1000 V/mm. This evaluation was performed with a CONV dose rate electron beamline in the A11-VAR ion chamber, with variable electrode spacing of 0.2–1.0 mm. Dashed lines indicate ±2%.

### 3.2.1 Dose per pulse response

The CCE and $P_{pol}$ values for the A11-VAR chamber, with 0.2–1.0 mm electrode spacing at an electric field strength of 1000 V/mm for DPP values of 1–9 Gy with corresponding PWs ranging from 0.5 to 4 µs, are shown in Figure 3. The CCE decreased with increasing DPP and increased PW at the same SSD. The CCE was also found to be higher under the same DPP for the A11-VAR ion chamber with smaller electrode spacings. Specifically, the CCE was found to decrease when the DPP was increased from 1 to 9 Gy for the A11-VAR ion chamber at a constant electric field strength of 1000 V/mm, with the 0.2-mm ion chamber having the highest CCE as a function of DPP (Figure 3A). The range of $P_{pol}$ values measured in a UHDR beam were comparable with the $P_{pol}$ values measured with a conventional dose rate beam for the 1.0 mm (0.997), the 0.6 mm (0.999), the 0.3 mm (0.993), and the 0.2 mm (0.992) configurations of the A11-VAR ion chambers (Figure 3B) (Table 3) regardless of the DPP.

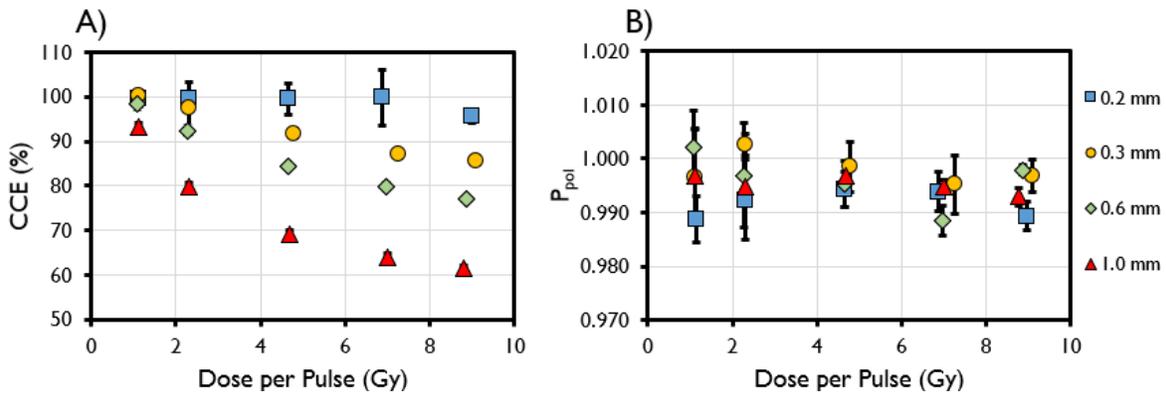

**Figure 3.** (A) Charge collection efficiency (CCE) and (B) polarity correction factor ($P_{pol}$) measured in the A11-VAR ion chamber, with variable electrode spacing ranging from 0.2 mm to 1.0 mm at doses per pulse of 1–9 Gy corresponding to pulse width settings of 0.5-4 µs.

TABLE 3. Measured $P_{ion}$, $P_{pol}$, and calculated average ion collection times measured in the tested ion chambers under conventional dose rate irradiations for the range of electrode spacing of 0.2-1.0 mm at 1000 V/mm (A11-VAR) or electrode spacing of 0.3 mm and electric field strengths of 500-1200 V/mm (A11-TPP and A30).

|  | Parallel Plate Ion Chambers Tested | | |
| --- | --- | --- | --- |
|  | A11-VAR | A11-TPP | A30 |
| $P_{ion}$ at tested electric field strengths and electrode spacing | 1.000–1.005 | 1.000–1.007 | 1.003–1.009 |
| $P_{pol}$ at tested electric field strengths and electrode spacing | 0.992–0.999 | 0.987–0.993 | 0.989–0.995 |
| Average ion collection time, μs, at tested electric field strengths and electrode spacing | 0.9–10.6 | 1.3–3.2 | 1.3–3.2 |

## 3. 2 | A11-TPP and A30 ion chambers

### 3.2.1 Electric field strength

Relative to the signal measured at 1000 V/mm, the mean signal was found to be within ±2% at electric field strengths of 333–1500 V/mm for the A11-TPP ion chamber and 333–2000 V/mm for the A30 ion chamber. At electric field strengths greater than 1500 V/mm for the A11-TPP and 2000 V/mm for the A30 ion chamber, exponential growth in signal per dose delivered at CONV dose rates was noted, indicating charge multiplication (Figure 4).

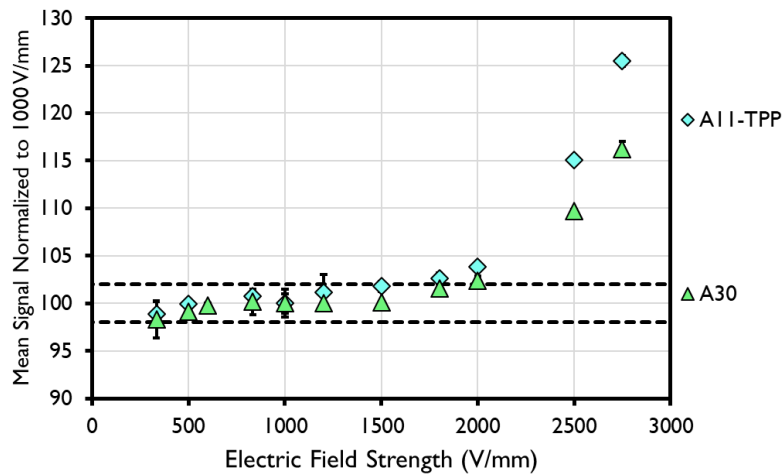

**Figure 4.** Mean signal plotted as a function of electric field strength (V/mm) normalized to the mean signal measured at 1000 V/mm. This evaluation was performed in a CONV dose rate electron beamline and tested in the A11-TPP and A30 ion chambers with electrode spacing of 0.3 mm. Dashed lines indicate ±2%.

The $P_{ion}$, $P_{pol}$, and average ion collection times measured in the A11-TPP and A30 ion chambers at 500–1200 V/mm are shown in Table 3.

### 3.2.2 Dose per pulse response

The dependence of the ion chamber response on DPP in terms of CCE and $P_{pol}$ at electric field strengths of 500–1200 V/mm was investigated at DPP values of 0.3–9 Gy at a single PW of 4 µs and using different SSDs. The CCE was shown to decrease as a function of DPP as the DPP increased from 0.3 to 9 Gy in the A30 and A11-TPP ion chambers (Figure 5A,B). The simulated CCE in a 0.3-mm-spaced ion chamber was found to decrease from ≥99% to 65% at 500 V/mm, from ≥99% to 89% at 1000 V/mm, and from ≥99% to 96% at 1200 V/mm (Figure 5C) at DPPs of 1–10 Gy. The measured $P_{pol}$ ranges for the A11-TPP chamber were 0.985–1.002 at an electric field strength of 500 V/mm, 0.992–1.015 at 100 V/mm, and 0.986–1.014 at 1200 V/mm (Figure S1A). The measured $P_{pol}$ ranges for the A30 chamber were 0.931–0.962 at an electric field strength of 500 V/mm, 0.939–0.959 at 1000 V/mm, and 0.947–0.965 at 1200 V/mm (Figure S1B). The simulated $P_{pol}$ values all exceeded 0.999 at electric field strengths of 500–1200 V/mm (Figure S1C) at DPPs of 1–10 Gy.

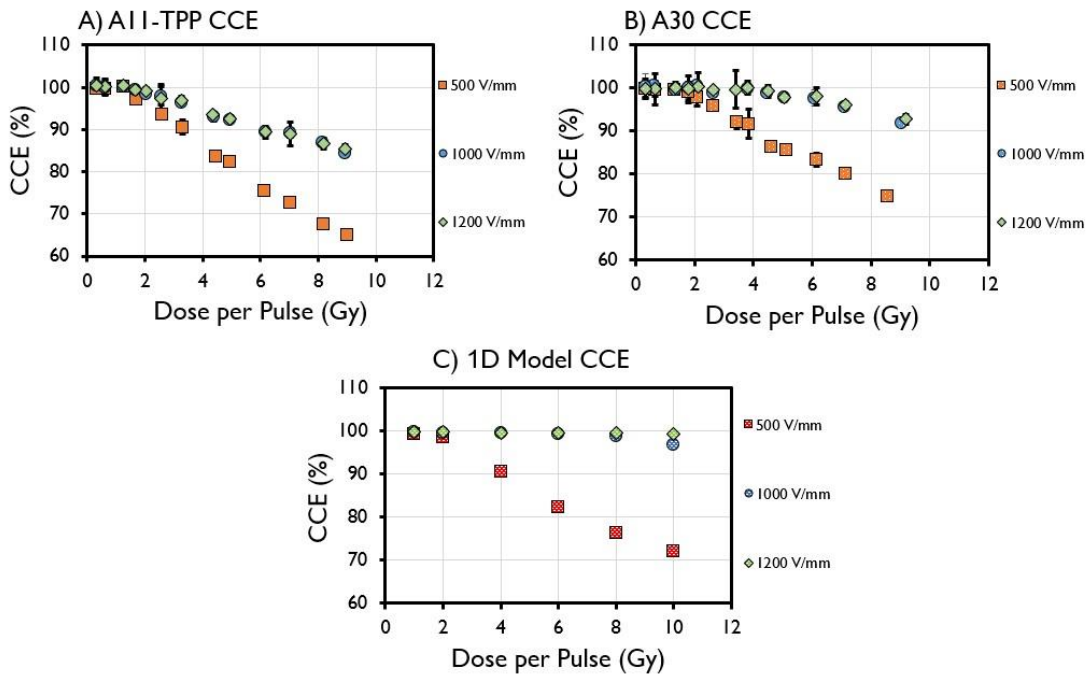

**Figure 5.** Charge collection efficiency (CCE) measured in the (A) A11-TPP, (B) A30, and (C) 1D model of ion chambers with 0.3-mm electrode spacing as a function of dose per pulse, with a pulse width setting of 4 µs. The electric field strengths evaluated were 500–1200 V/mm.

### 3.3 | Mean dose rate response

The CCE and $P_{pol}$ values were found not to be dependent on PRFs within the tested range of 10

Hz to 120 Hz (<1% relative standard deviation in the measured CCE and $P_{pol}$ values as a function of PRF at a given electric field strength). The measured CCE values in each ion chambers ranged from 65% to 85% for the A11-TPP chamber and from 75% to 93% for the A30 chamber at electric field strengths of 500–1200 V/mm, with constant CCE as a function of PRF at a given electric field strength. The measured $P_{pol}$ values in each ion chamber were 0.995–1.020 for the A11-TPP chamber and 0.905–0.955 for the A30 chamber at electric field strengths of 500–1200 V/mm for PRFs of 10-120 Hz (Figure 6).

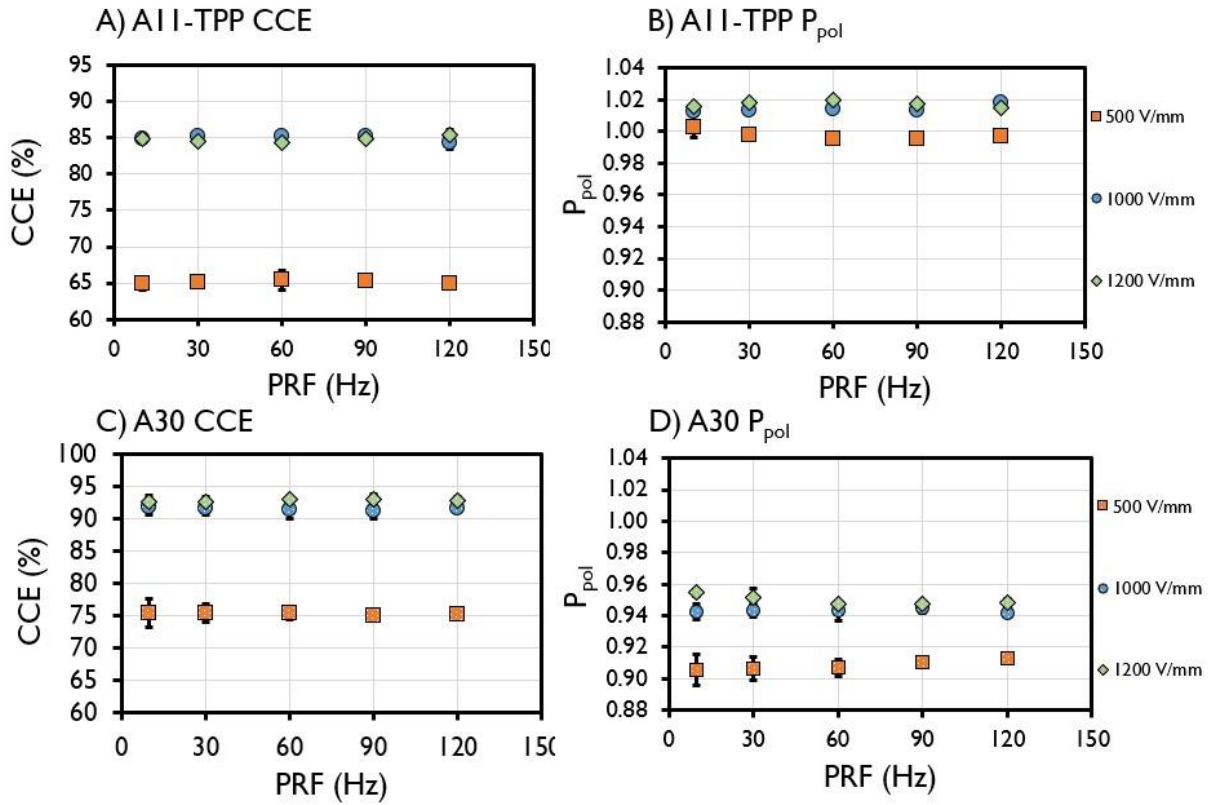

**Figure 6.** Charge collection efficiency (CCE) and polarity correction factor ($P_{pol}$) measured in the (A,B) A11-TPP and (C,D) A30 ion chambers with 0.3-mm electrode spacing at pulse repetition frequencies (PRF) of 10–120 Hz at a dose per pulse of 9 Gy and pulse width setting of 4 μs. The electric field strengths evaluated were 500–1200 V/mm.

### 3.4 | Instantaneous dose rate (pulse width) response at a constant dose per pulse

In both the A11-TPP and A30 ion chambers, the CCE measured at a DPP of 1 Gy was found to have a relative standard deviation of 1% for PWs ranging from 0.5 μs to 4 μs at electric field strengths of 500 V/mm and 1000 V/mm. When the DPP was 2 Gy and the electric field strength was 500 V/mm, the CCE depended on the PW in both chambers for the smaller PWs, with differences in CCE of up to 5% (for the A11-TPP) and 3%-5% difference (for the A30) at the shortest PW investigated for both chambers. When the electric field strength was 500 V/mm, the CCE depended on the PW in the A11-TPP and A30 chambers at DPPs of 2 Gy or higher for PWs as low as 2 µs (Figure 7). At a field strength of 1000 V/mm, this dependence disappeared and the CCE across the settings investigated was found to be within 1% of each other (relative standard deviation).

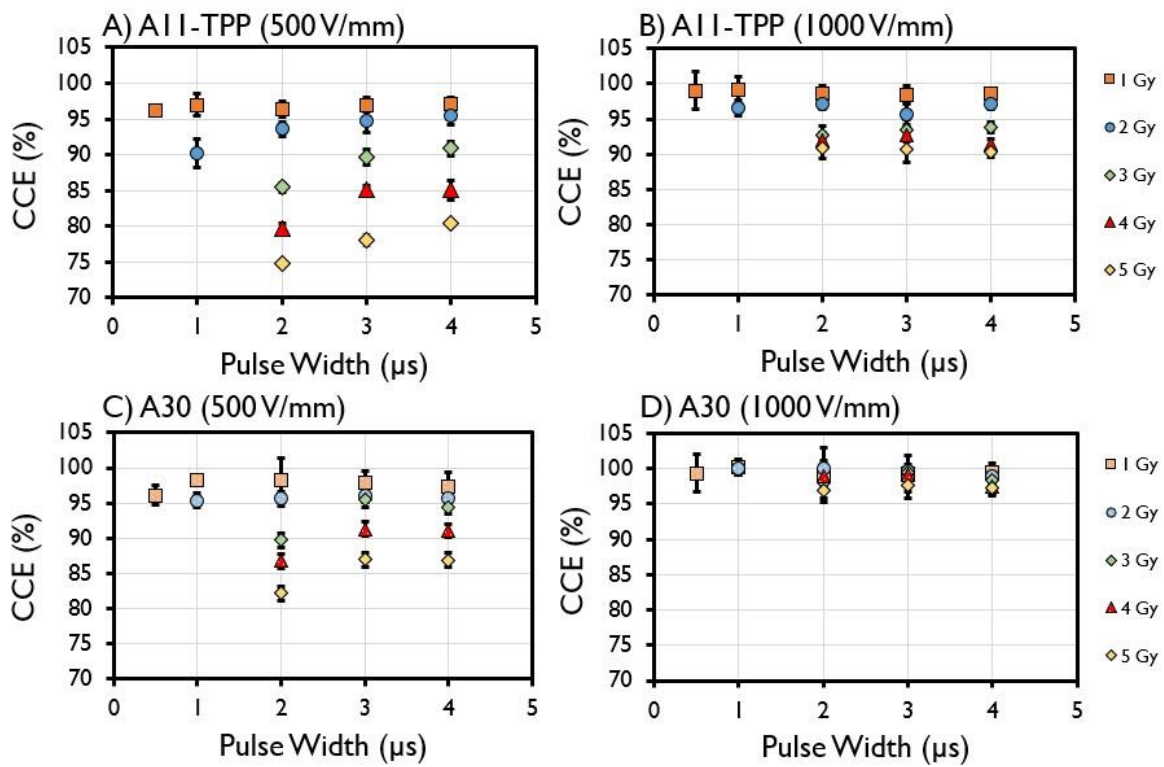

**Figure 7.** Charge collection efficiency (CCE) measured in the A11-TPP and A30 ion chambers at doses per pulse of 1–5 Gy for matched pulse width settings of 0.5–4 µs. The electric field strengths evaluated were 500 and 1000 V/mm.

### 3.5 | Simulation of dose per pulse and pulse width dependence

The simulated CCE for identical DPP deliveries (1–10 Gy) at different PWs (0.5–4 µs), with a fixed electrode spacing of 0.3 mm and electric field strengths of 500–1200 V/mm, are shown in Figure

8. Comparing the CCE value generated by the simulations (Figure 8) with the measured values (Figure 7) revealed that the measured values in the A30 and A11-TPP ion chambers were lower than those generated by the simulations. However, the simulations did match the measurements in that at 1000 V/mm, no PW dependence (<1%) was found at DPPs of 1 Gy (PW 0.5–4 µs), 2 Gy (PW 1–4 µs), or 3–5 Gy (PW 2–4 µs). At 500 V/mm, CCE did not depend on PW at DPPs as low as 2 Gy and PWs as low as 1 µs. At 1200 V/mm, compared with the CCE at 1000 V/mm, the CCE value showed marginal improvement at the measured DPP values (1–5 Gy) and less of a PW dependence at 0.5 µs. However, at the much higher DPP settings of 10 Gy, the CCE was shown to improve by 2.5 to 6.3%, especially in the lower PW settings.

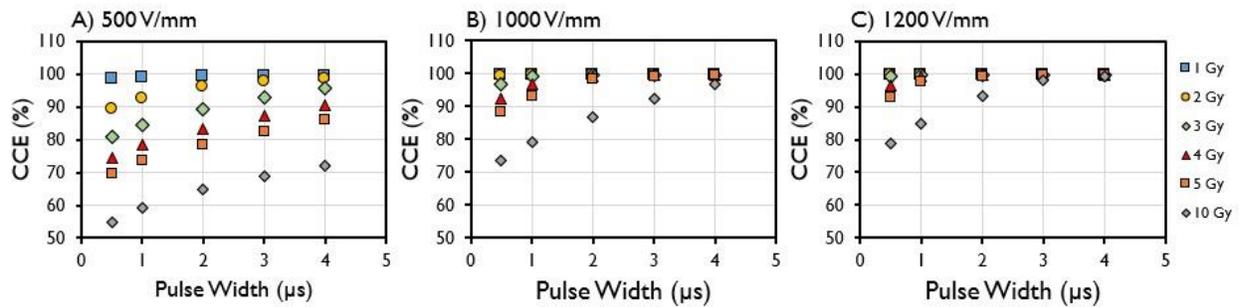

**Figure 8.** Charge collection efficiency (CCE) simulated in a 1D model ion chamber with 0.3-mm electrode spacing at electric field strengths of (A) 500 V/mm, (B) 1000 V/mm, and (C) 1200 V/mm at doses per pulse of 1–10 Gy as a function of pulse widths of 0.5–4 µs.

The measured CCE for identical DPP deliveries (1–10 Gy) at a fixed electric field strength of 1000 V/mm and different PW (0.5–4 µs) and electrode spacings of 0.1–1.0 mm are shown in Figure 9. At an electrode spacing of 0.1 mm, CCE did not depend on PW at DPPs up to 10 Gy. However, for ion chambers with electrode spacings of 0.3 mm and higher, the dependence of CCE on PW was exacerbated at lower PWs, higher DPPs, and larger electrode spacings.

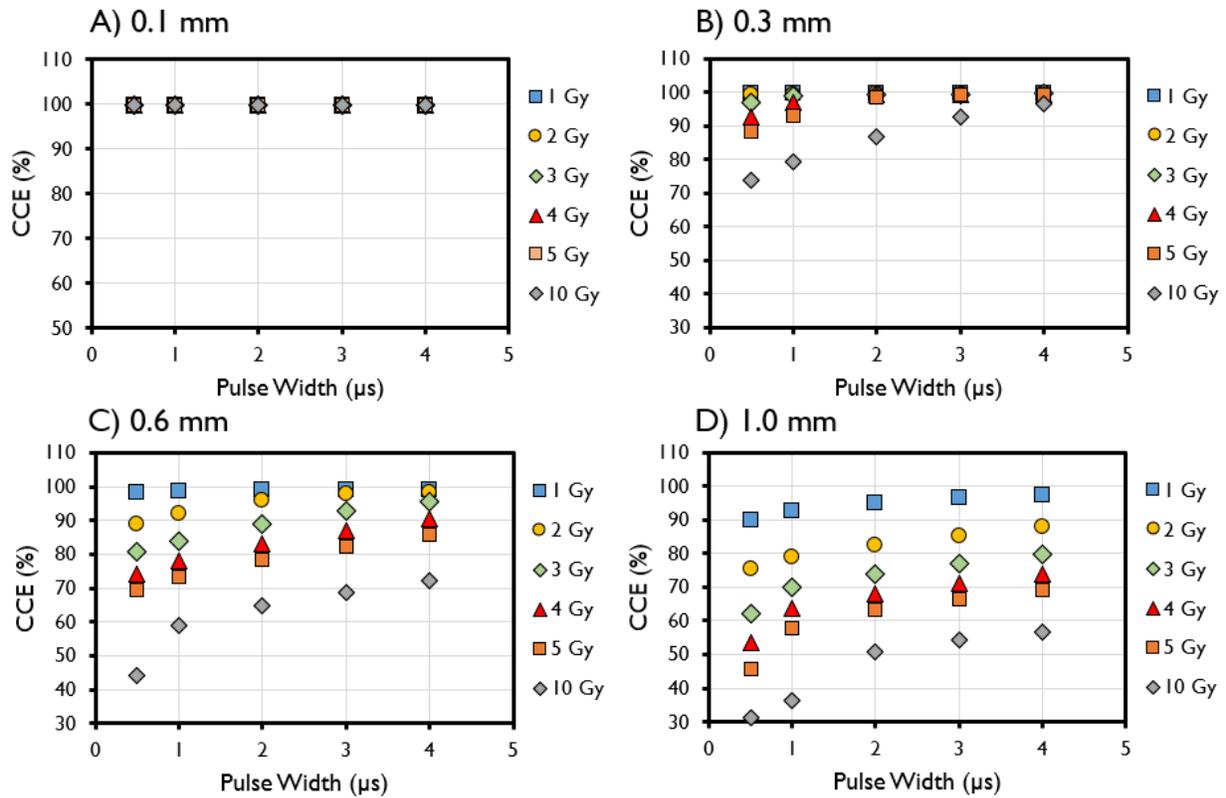

**Figure 9.** Charge collection efficiency (CCE) measured in a 1D model ion chamber at electric field strength of 1000 V/mm for electrode spacings of (A) 0.1 mm, (B) 0.3 mm, (C) 0.6 mm, and (D) 1.0 mm at doses per pulse of 1–10 Gy.

## 4 | DISCUSSION

Here we tested the performance of ultra-thin parallel plate ion chambers with sub-mm electrode spacing. Although ion chambers in general are rarely used at electric field gradients set at or above 1000 V/mm for CONV dose rate beamlines, we showed that under UHDR conditions, such chambers are necessary and that the measured charges per dose were within 1%–2% of the charge collected at electric field strengths used typically in ion chambers (up to 1500 V/mm for the A11-TPP and up to 2000 V/mm for the A30). At and above 1500 V/mm, the collected charge measured per dose shifted towards the proportional counting region, as demonstrated by the exponential increase in the charge collected indicating a charge multiplication effect,[55] although others have found the charge multiplication region to be at 2000 V/mm under conventional dose rate conditions.[56,57] The charge multiplication effect under UHDR conditions was demonstrated by Kranzer et al, who reported a CCE of >100% under UHDR conditions for a parallel plate chamber with 0.25-mm electrode spacing operating at an electric field strength of 2000 V/mm.[39]

When evaluating chamber response at different electric field strengths under UHDR conditions, we showed that ion chambers operating at electric field strengths of 1000 V/mm yielded higher CCE than chambers operating at 500 V/mm under identical UHDR conditions, whereas the chamber response at 1200 V/mm provided negligible improvement in CCE, compared with 1000 V/mm, at the range of DPP values evaluated. However, an improvement in CCE at electric field strengths greater than 1000 V/mm might be observed under higher DPP conditions (DPP > 9 Gy), as indicated by the simulations performed, but was not experimentally tested in this study because of limitations in machine output. Given the limited improvements in performance of the ion chambers tested and to reduce the risk of potential charge multiplication, we recommend using these chambers at a field strength of 1000 V/mm for DPPs <10 Gy.

In evaluating the A11-VAR, we showed experimentally that the CCE improves with decreasing electrode spacing despite operating at the same electric field strength, with the 0.2 mm configuration showing ≥99% CCE at a DPP up to 7 Gy and 96% CCE at a DPP of 9 Gy, thus demonstrating that smaller electrode spacings are more critical for ion chamber performance in UHDR conditions than electric field strength, likely because the charge carriers traversed a shorter distance to reach the electrodes. However, we found that the experimental performance of all ion chambers fell short of the values generated from their respective simulations. This may have been caused by limitations in the current design of the ion chamber prototypes or in electrometer design thereby causing saturation in the signal collected, which would lower the CCE measured as a function of DPP, as well as the assumptions that underlie the simulation model used. The SuperMax has a sampling rate of 10 Hz and is limited by a current of 500 nA before saturating, so the maximum charge it would be able to measure in a 10 Hz sample would be at 50 nC. The DPP needed to reach the 50 nC threshold would depend on the volume of the ion chamber used as well as the electric field strength. We found experimentally that for the A11-VAR, $P_{pol}$ did not depend on electrode spacing (0.2–1.0 mm) or DPP (1–9 Gy) at the same electric field strength of 1000 V/mm (Figure 3). Other published studies have shown a polarity effect in chamber response as a function of varying the DPP, but this may have been due to the relatively low electric field strengths that were tested in parallel plate chambers (<500 V/mm) that had electrode spacings of 1 mm or to stem effects that arise from changing the SSD in DPP evaluations.[37,40,50]

       The A11-VAR ion chamber was used for information on optimal electrode separation based on the following criteria: (1) CCE, (2) electric field strength, and (3) $P_{pol}$. The consensus based on the data generated was that ion chambers with smaller electrode spacings and higher electric field strengths perform better in UHDR conditions, but when considering limitations in current manufacturing techniques of ion chambers, electrode spacings smaller than 0.3 mm becomes limiting in terms of reproducibility. Based on these criteria and the data generated with the A11-VAR, two new prototype chambers were produced with a fixed 0.3-mm electrode separation but with different collecting electrode diameters to evaluate potential volume effects in UHDR conditions. We found that the A30, which has a smaller collection volume and diameter than the A11-TPP, yields improved performance in CCE under UHDR conditions but impaired $P_{pol}$ performance. We found that $P_{pol}$ under UHDR conditions for both the A11-TPP and the A30 did not significantly depend on electric field strength, DPP, PRF, or PW (as confirmed by experimental data and simulations). Rather, differences in their responses were likely an effect of their inherent construction design. The $P_{pol}$ response seen for these parallel-plate chambers are in stark contrast to what has been observed for a cylindrical chamber under UHDR conditions, where $P_{pol}$ for the A26 microchamber was strongly affected by DPP, but not for PRF or PW, likely owing to voltage-dependent polarity effects of this chamber as observed in conventional dose rate conditions.[42,58] Several studies have shown that even at conventional dose rate beamlines, ion chambers with small sensitive volumes are more susceptible to the polarity effect because of greater distortions in the electric field lines from the small potential difference between the guard and inner collecting electrode and residual positive space charge effects formed when the liberated electrons reach the positive electrode without forming negative ions.[59-64] However, one limitation in the current study is that we did not test the $P_{pol}$ dependence under UHDR conditions at different beam energies, though Petersson et al reported that such a polarity effect becomes magnified under lower energy electron beams (5–6 MeV)[37].

Previous studies have suggested that ion recombination and the CCE depend primarily on the polarizing voltage and the DPP.[36,37] However, in evaluating the instantaneous dose rate or PW dependence in this study, at electric field strengths of 1000 and 1200 V/mm, no significant difference in the CCE as a function of PW at a given DPP was found. However, at an electric field strength of 500 V/mm, a PW dependence was noted in the experimental data indicating that 500 V/mm was insufficient to adequately collect the charge carriers below certain PWs at a given DPP. Specifically, at a DPP of ≥2 Gy, the CCE was lower for shorter PWs at 500 V/mm. The PW dependence is exacerbated at higher DPPs and may be a feature of the temporal and spatial charge carrier imbalance that is generated within the collecting volume, where the free electrons approach the positive electrodes significantly faster than positive ions as reported by Kranzer et al[39]. The reason is that at shorter PWs, the same dose is delivered within a shorter amount of time within a pulse, which would allow greater electric field imbalance, charge recombination, and lower CCE. Potential methods for mitigating this PW dependence include using higher electric field strengths and smaller electrode spacings, as the collected charge would travel shorter distances with greater force applied to reach the electrodes (as confirmed by the data from simulations in Figures 8 and 9). Thus, the data collected support the statement that the PW dependence of ultra-thin ion chambers in UHDR beamlines depends on a combination of the DPP, PW, electrode spacing, and electric field strength, with the CCE being lower at smaller PWs, smaller electric field strengths, larger DPPs, and larger electrode spacings.

## 5 | CONCLUSION

In this work, we evaluated the CCE and polarity response of three distinct ultra-thin ion chambers under UHDR and UHDPP conditions with DPP values of up to 9 Gy and mean dose rates of up to 1620 Gy/s. We demonstrated that the CCE depends strongly on the electrode spacing, electric field strength, the PW, and the DPP of the beam delivered, and that the CCE was independent of other beam parameters such as total dose, mean dose rate, and instantaneous dose rate. A CCE of ≥ 95% was achieved with the A30 chamber for DPP up to 5 Gy with no observable dependency on PW, while still achieving an acceptable performance in conventional dose rate beams, opening up the possibility for this type of chamber to be used as a reference class chamber for calibration purposes of electron FLASH beamlines.

## ACKNOWLEDGEMENTS

We thank Christine F. Wogan, MS, ELS, of MD Anderson's Division of Radiation Oncology for editorial contributions to this article. Research reported in this publication was supported by the National Cancer Institute of the National Institutes of Health under Topic 434-75N91022C00039 - Phase I SBIR Contract with Standard Imaging Inc., R01CA266673, and P30 CA016672; by the University Cancer Foundation via the Institutional Research Grant program at MD Anderson Cancer Center; by the University of Texas MD Anderson Cancer Center UTHealth Graduate School of Biomedical Sciences Dr. John J. Kopchick Fellowship, Houston, TX ; by the University of Texas MD Anderson Cancer Center UTHealth Graduate School of Biomedical Sciences American Legion Auxiliary Fellowship, Houston, TX ; and by UTHealth Innovation for Cancer Prevention Research Training Program Predoctoral Fellowship (Cancer Prevention and Research Institute of Texas grant RP210042). The content is solely the responsibility of the authors and does not necessarily represent the official views of the National Institutes of Health or the Cancer Prevention and Research Institute of Texas.

## CONFLICT OF INTEREST

Shannon Holmes, Larry DeWerd, and Brian Hooten are employees at Standard Imaging.

Supplementary Data

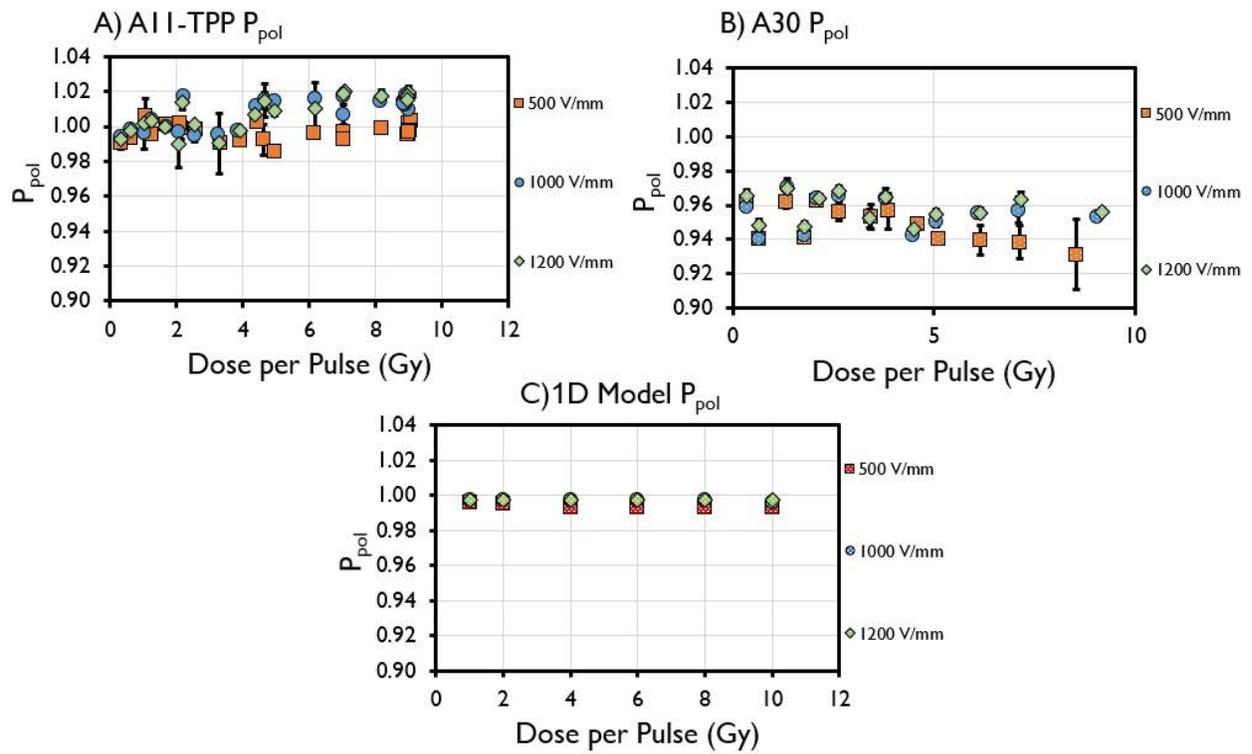

**Figure S1.** $P_{pol}$ measured in the (A) A11-TPP, (B) A30, and (C) 1D model of ion chambers with 0.3-mm electrode spacing as a function of dose per pulse at a pulse width setting of 4 μs. The electric field strengths evaluated were 500–1200 V/mm.